\documentclass{JHEP3}
\usepackage{epsfig}
\def\AdSs5{$AdS_5$}
\def\AdS5s5{$AdS_5 \times S^5$}

\def\calN{{\cal N}}

\def\calT{{\cal T}}

\def\sign{{\rm sign}}
\def\p+{{p_-}}

\def\hQ{{Q^-}}

\def\half{\frac{1}{2}}

\newcommand{\ie}{{\it i.e.}}

\newcommand{\be}{\begin{equation}}
\newcommand{\ee}{\end{equation}}
\newcommand{\ba}{\begin{eqnarray}}
\newcommand{\ea}{\end{eqnarray}}
\newcommand{\bea}{\begin{eqnarray}}
\newcommand{\eea}{\end{eqnarray}}
\newcommand{\nn}{\nonumber}

\def\tS{\tilde S}

\def\bbbone {{\mathchoice {\rm 1\mskip-4mu l} {\rm 1\mskip-4mu l}
{\rm 1\mskip-4.5mu l} {\rm 1\mskip-5mu l}}}
\def\Zop{\rlx\leavevmode\ifmmode\mathchoice{\hbox{\cmss Z\kern-.4em Z}}
 {\hbox{\cmss Z\kern-.4em Z}}{\lower.9pt\hbox{\cmsss Z\kern-.36em Z}}
 {\lower1.2pt\hbox{\cmsss Z\kern-.36em Z}}\else{\cmss Z\kern-.4em
 Z}\fi}
\def\hQ{Q^-}
\def\bbbone {{\mathchoice {\rm 1\mskip-4mu l} {\rm 1\mskip-4mu l}
{\rm 1\mskip-4.5mu l} {\rm 1\mskip-5mu l}}}

\newcommand{\bfa}{\mbox{\textit{\textbf{a}}}}
\newcommand{\bfz}{\mbox{\textit{\textbf{z}}}}

\title{Non-perturbative contributions to  the plane-wave string mass matrix}
\author{Michael B. Green$^\dagger$, Stefano Kovacs$^{\#}$ and
  Aninda Sinha$^\dagger$ \\ {$^\dagger$ \it   Department of Applied
Mathematics and
Theoretical Physics,\\ Wilberforce Road, Cambridge CB3 0WA,
UK} \\ \email{M.B.Green, A.Sinha@damtp.cam.ac.uk} \\
$^{\#}$ Max-Planck-Institut f\"ur Gravitationsphysik \\
Albert-Einstein-Institut \\
Am M\"uhlenberg 1, D-14476 Golm, Germany  \\
\email{stefano.kovacs@aei.mpg.de}}

\abstract{$D$-instanton contributions to the mass matrix of arbitrary
excited string states of type IIB
string
theory in the maximally supersymmetric plane-wave background are calculated to leading order in the string coupling using a
supersymmetric light-cone boundary state formalism.  The explicit
non-perturbative dependence of the mass matrix on the complex string
coupling, the plane-wave mass parameter and the mode numbers of the
excited states is determined. }

\preprint{DAMTP-2005-22 \\ AEI-2004-058}

\keywords{D-instanton, plane-wave, AdS/CFT}

\begin{document}

\section{Introduction}
\label{intro}

The conjectured correspondence between the BMN sector of ${\cal N}=4,
d=4$ supersymmetric
Yang-Mills \cite{bmn} and type IIB string theory in the maximally
supersymmetric plane-wave background \cite{mt} has been examined in some
detail at the
perturbative level \cite{reviews,gutjahr,recentgauge}. However, the
understanding of non-perturbative
aspects of the correspondence has been very limited
(although $D$-branes were constructed on the string
theory side in \cite{dbranes,gg}).
Such non-perturbative effects are well-studied
in the context of the AdS/CFT
correspondence where Yang--Mills instanton effects in $\calN =4$
supersymmetric
Yang-Mills correspond closely to $D$-instanton effects in type IIB
superstring theory  in $AdS_5\times S^5$ \cite{bgkr}-\cite{ymrev}. A
natural
question to ask is whether there is a similar relationship between
non-perturbative effects
in plane-wave string theory and the BMN limit of the gauge theory.

According to the correspondence between plane-wave string theory and the
BMN limit of
$\calN =4$ Yang--Mills theory,  the light-cone gauge string theory mass
matrix is related to
the gauge theory conformal dimension.  More precisely, the conformal
dimension, $\Delta$,
and  $R$-charge, $J$, are given by
\be
{H^{(2)}\over \mu}={\Delta-J}\,,
\ee
where $H^{(2)}$ denotes the two-particle hamiltonian
and $\mu$ is the constant background $RR$ (Ramond--Ramond) five-form flux.
 The   two-particle hamiltonian is the sum of two pieces
\be
H^{(2)}=H^{(2)}_{\rm pert}+H^{(2)}_{\rm nonpert}\,.
\ee
The perturbative part, $H^{(2)}_{\rm pert}$, is a power series in the
string
coupling,  $g_s$.
These perturbative contributions to the mass spectrum have been
analysed in some detail on both the string side and in the BMN limit
of the gauge theory using the technology developed in \cite{pert}.
 There have been no calculations of
non-perturbative corrections due to $D$-instanton effects, which
contribute to  $H^{(2)}_{\rm nonpert}$. It is the purpose of this paper to
fill the gap in the existing literature and open the possibility of
examining the proposed BMN/plane-wave correspondence at the
non-perturbative level.

The single $D$-instanton sector  has a measure that is proportional to
\be
e^{2i\pi \tau}\, g_s^{7/2}\, ,
\label{dmeasure}
\ee
where $\tau= \tau_1 + i \tau_2 \equiv C^{(0)} + i e^{-\phi}$
($C^{(0)}$ is the Ramond--Ramond pseudoscalar, $\phi$ is the dilaton
and $g_s= e^{\phi}$).  The factor $g_s^{7/2}$ can be extracted from
the form of certain higher derivative interactions that enter into the
type IIB effective action at
$O(1/\alpha')$ \cite{mbgrev}{\footnote{More precisely, the complete
    dilaton dependence of any such process is a modular form that
    contains specific multi D-instanton contributions, from which one
    can read off the measure, including the $g_s^{7/2}$ factor.}}.
Although (\ref{dmeasure})  is exponentially small, it is the
leading contribution  with the phase factor $e^{2\pi i C^{(0)}}$.  It
is therefore of  interest to understand how the mass matrix is
modified by these contributions.  In the following we will calculate
such $D$-instanton contributions to mass matrix elements to leading
order in the string coupling.  These can be compared with the gauge
theory instanton contributions to the corresponding two-point
functions  in the BMN limit, which will be the subject of a separate
work \cite{gks}.

We will use the light-cone boundary state description of the
$D$-instanton \cite{gg,ggut} to evaluate elements of the mass matrix for
arbitrary string states.  Equivalently, the leading contribution to
the two-point function of string states in the $D$-instanton
background will be associated with a world-sheet that is the product
of two disks, with one closed-string state attached to each, and with
Dirichlet boundary conditions. These boundary conditions impose the
condition that the $D$-instanton has a position given by the bosonic
moduli $x_0^I$, $x_0^+$, $x_0^-$.  The sixteen broken kinematical and
dynamical supersymmetries lead to the presence of sixteen fermionic
moduli, $\epsilon^a$, $\eta^{\dot a}$, which are included by attaching
a total of sixteen fermionic open-string states to the boundaries of
the
disks. All the moduli are then integrated.
This procedure will be implemented by use of the boundary state formalism.

This paper is organised as follows.  Some of the notation and
conventions of free light-cone plane-wave string theory are  reviewed
in section 2.1 and the structure of the $D$-instanton boundary state
is reviewed in section 2.2.  In section 2.3 the dependence on the
fermionic moduli is obtained by applying the eight broken kinematical
supersymmetries and the eight broken dynamical supersymmetries to the
boundary state.  This results in a `dressed'  boundary state.  In
order to evaluate the $D$-instanton contribution to two-point
functions of string states this dressed boundary state is generalised
to a composite two-string boundary state that is the product of two
single string boundary states.  Matrix elements of this state with
physical two-string states give the $D$-instanton
contribution to the elements of the mass matrix to leading order in the string
coupling.  Equivalently, we are evaluating the product of two disks
with a single physical closed-string state attached to each and a
total of sixteen open fermionic strings (representing the fermionic
moduli) attached to the disk boundaries, which satisfy Dirichlet
boundary conditions in all space-time directions.  In section 2.3.1 we
will see that there are no $D$-instanton corrections to the mass
matrix with external supergravity states, as expected. The
contributions of a single $D$-instanton to the mass matrix of massive
string states created by the action of string creation modes are
considered in sections 3 and 4.  In section 3, states with  two
oscillator excitations (two impurities) are considered in detail.  We
will see that the leading contributions to $H^{(2)}_{\rm nonpert}/\mu$ behave as
$g_s^{7/2} m^3$ when $m$ is large, independent of the mode number of
the oscillators.  States with four bosonic string oscillator
excitations are considered in section 4.  The elements of the mass
matrix between such states behave as $g_s^{7/2} m^7/r^2 s^2$, where
$r$, $-r$, $s$, $-s$  are the mode numbers of the oscillators on each
of the strings\footnote{These arise in pairs of equal magnitude and
opposite sign. They are also equal on the two strings as a result of
the conservation of light-cone energy.}.   States
with fermionic excitations and states with larger numbers of
impurities will also be discussed in section 4.  We conclude with a
summary and discussion of the implications for the BMN limit of $\calN
=4$ supersymmetric Yang--Mills theory in section 5.

\section{Couplings of the plane-wave $D$-instanton}
\subsection{Plane-wave string theory}

In this subsection and the appendix we will review some properties of
free plane-wave light-cone gauge string theory \cite{mt}.  The
expressions for the supercharges and the hamiltonian are given in the
appendix in terms of integrals over the usual superstring world-sheet
fields, $x^I$, $S$ and $\tilde S$.  Here we will summarise the mode
expansions of these expressions.

The free quantum mechanics hamiltonian \cite{mt} is given
by\footnote{We here use the lower case symbol $h$ to distinguish the first
quantised hamiltonian from the lowest order contribution to the
string field theory hamiltonian.}
\ba
2p_-\,  h&=& m \left(a^{\dagger\, I} a^I+i S_0^a \Pi_{ab} \tilde
S_0^b+4\right) +\sum_{k=1}^\infty\left[\alpha^I_{-k}\alpha_k^I
+\tilde\alpha^I_{-k}\tilde\alpha_k^I+
\omega_k\left(S^a_{-k}S^a_{k}+\tilde
S^a_{-k}\tilde S^a_k\right)\right]\nonumber \\
&=&m\left(a^{\dagger\, I} a^I+\theta_L^a\bar
\theta_L^a+\bar\theta_R^a\theta_R^a\right)+\sum_{k=1}^\infty
\left[\alpha^I_{-k}\alpha_k^I+\tilde\alpha^I_{-k}\tilde\alpha_k^I
+\omega_k\left(S^a_{-k}S^a_{k}+
\tilde    S^a_{-k}\tilde S^a_k\right)\right]\,,\nonumber \\
\label{hamilstring}
\ea
where $m=\mu\, p_-\,\alpha'$, $\omega_n={\rm sign}(n)\sqrt{m^2+n^2}$.  The
matrix $\Pi$ is defined in terms of $SO(8)$ $\gamma$ matrices by
\be
\Pi = \gamma^1\gamma^2\gamma^3\gamma^4\, ,
\label{pidef}
\ee
and its presence in (\ref{hamilstring}) implies that $h$ is invariant
under $SO(4)\times SO(4)$ rather than $SO(8)$.  The choice of the
matrix $\Pi$ in (\ref{pidef}) reflects the choice of specific
directions for the constant $RR$ five-form flux which
defines the background.

The modes of the left-moving and right-moving bosonic and fermionic
world-sheet fields, $\alpha$, $\tilde\alpha$, $S$ and $\tilde S$,
satisfy the (anti)commutation relations,
\bea
[\alpha_m,\alpha_n]&=&\omega_n \delta_{m+n}\,,\quad [\tilde
  \alpha_m,\tilde\alpha_n]=\omega_n \delta_{m+n}\, ,\nn\\
\{S^a_m,S^b_n\}&=&\delta^{ab}\delta_{m+n}\,,\quad \{\tilde S^a_m,\tilde
S^b_n\}=\delta^{ab}\delta_{m+n}\,,
\label{commrels}
\eea
while all other (anti)commutators of these variables vanish.
The zero-mode fermionic variables $\theta_R $, $\bar\theta_R $, $\theta_L $ and
$\bar\theta_L$ in (\ref{hamilstring}) are  four-component spinors with
$SO(4)\times SO(4)$ chiralities $(+,+)$ and $(-,-)$, defined in terms of
$S_0^a$ and $\tilde S_0^a$ by
\bea
\theta_R &=&\frac{1}{2\sqrt 2}(1+\Pi) \, (S_0^a + i \tilde S_0^a)\, ,
\qquad \bar \theta_R =\frac{1}{2\sqrt 2}(1+\Pi) \, (S_0^a -
i \tilde S_0^a) \,  , \nn\\
\theta_L  &=&\frac{1}{2\sqrt 2}(1-\Pi) \, (S_0^a + i \tilde S_0^a)\, ,
\qquad \bar \theta_L =\frac{1}{2\sqrt 2}(1-\Pi) \, (S_0^a -
i \tilde S_0^a) \, .
\label{fermdefs}
\eea

The non-zero anticommutation relations between these spinors are
\be
\{\bar\theta_R, \theta_R\} = \frac{(1+\Pi)}{2}\, ,
\qquad\{\bar\theta_L, \theta_L\}=\frac{(1-\Pi)}{2}\,.
\label{anticomm}
\ee

The transverse position and momentum operators pair together to form
harmonic oscillator creation and annihilation operators,
\be
{a^{\dagger\, I}}={1\over \sqrt{2|m|}}(p_0^I+i|m|x_0^I),\quad { a}^I=
{1\over \sqrt{2|m|}}(p_0^I-i|m|x_0^I),
\ee
satisfying
\be
[{ a}^I,{a^{\dagger\, J}}]=\delta^{IJ}\,.
\ee
 The operators
${a}$ and ${ a^{\dagger}}$
are referred to as the zero mode bosonic oscillators.
The presence of the fermion mass-term in the hamiltonian explicitly breaks the
$SO(8)$ symmetry to $SO(4)\times SO(4)$.

Let us briefly review the massless sector of the theory.  Recall that
it is usual to take the BMN vacuum state,  $|0\rangle_h$, to be the
bottom state of a `massless' supermultiplet.  It is natural to use a
Fock space description for the fermions based on  $|0\rangle_{h}$ as
the ground state satisfying
\be
 \bar\theta_L|0\rangle_{h} = 0 =  \theta_R|0\rangle_{h}\, .
\label{fermspace}
\ee
In this basis the operators $\theta_L$ and $\bar\theta_R$ are creation
operators and are used to create the other states in the
multiplet.
The state $|0\rangle_{h}$ is a nondegenerate bosonic state
with $p_+ =0$.  All the other states have positive $p_+$ with equal
numbers of degenerate bosons and fermions.  The lowest lying levels of
the string are generated by acting with the zero bosonic and fermionic
modes on the ground state.  This generates towers of supergravity
states that include infinite numbers of Kaluza--Klein-like
excitations.  It will prove convenient in the following to refer to a
different basis in which the ground state is the complex dilaton,
$|0\rangle_D$, that is annihilated by $\theta_L$ and $\theta_R$.
Acting on this state with $\bar\theta_R$ and $\bar \theta_L$ generates
the 256 BPS states in a short supermultiplet. For example, acting with
four powers of $\bar\theta_L$ gives the BMN ground state,
\be
{1\over 4!}\epsilon_{a_1a_2a_3a_4}\bar\theta_L^{a_1}\bar\theta_L^{a_2}
\bar\theta_L^{a_3}\bar\theta_L^{a_4} |0\rangle_D = |0\rangle_{h}\, ,
\label{bmnstate}
\ee
where the $0$ indicates that the state is the ground state of the zero
mode bosonic harmonic oscillators. Similarly the conjugate of the BMN
ground state is obtained by acting with $\bar\theta_R$,
\be
{1\over 4!}\epsilon_{a_1a_2a_3a_4}\bar\theta_R^{a_1}
\bar\theta_R^{a_2}\bar\theta_R^{a_3}
\bar\theta_R^{a_4}|0\rangle_D
=  |0\rangle_{\bar h} \,.
\label{boundbmnbar}
\ee

In making contact with the boundary state description of the $D$-instanton
we will often consider matrix elements between dilaton ground states,
$|0\rangle_D$.  In
particular, the only non-zero matrix element in the space of the zero-mode
fermions (the $\theta$'s) is
\be
{}_{\bar D}\langle 0|0 \rangle_D\ =
{}_D\langle 0| \prod_{a=1}^4(\bar\theta_L^a\bar\theta_R^a) |0 \rangle_D\,.
\label{nondil}
\ee

We will  use a convention in which $p_- >0$ for incoming states and
$p_- <0$ for outgoing states.  After integration over
the instanton modulus $x_0^-$, $p_-$ is conserved, which means that for any
process involving $M$ incoming and $N$ outgoing states
\be
\sum_{r=1}^M p_{r-} + \sum_{s=1}^N p_{s-}=0\,.
\label{ppluscons}
\ee

The background preserves 32 supersymmetries. Sixteen of these are
kinematical and do not commute with the hamiltonian, while sixteen are
dynamical and commute with the hamiltonian.  The kinematical
supersymmetry generators are proportional to the $\theta$'s and $\bar
\theta$'s. They will be denoted by $q$ and $\bar q$, defined by
\bea
q_R = e(p_-)\sqrt{|p_-|}\theta_R, &\qquad& q_L =
e(p_-)\sqrt{|p_-|}\theta_L\nn\\
\bar q_R = \sqrt{|p_-|}\bar\theta_R, &\qquad& \bar q_L =\sqrt{|p_-|} \bar
\theta_L\, ,
\label{kinsym}
\eea
where $e(p_-)=\sign(p_-)$. The generators $q$ and $\bar q$
satisfy the standard anti-commutation relations $\{q, \bar q\} = p_-$.

The dynamical supersymmetry generators,  $Q$ and ${\tilde Q}$,
are given by
\begin{eqnarray}
\label{q-}
\sqrt{ 2 |p_-|}\, Q_{\dot a}
&=&  p_0^I \gamma^I_{\dot a b} S_0^b
   - |m| x_0^I \left(\gamma^I\Pi\right)_{\dot a b}
  {\tilde S}_0^b
\nonumber\\
&+& \sum_{n=1}^\infty
\left( c_n \gamma^{I}_{\dot a b}
       (\alpha_{-n}^I S_n^b + \alpha_n^I S^{b}_{-n} )
+
 \frac{{\rm i} m}{2\omega_n c_n }
 \left(\gamma^I\Pi\right)_{\dot a b}
       ({\tilde \alpha}_{-n}^I {\tilde S}_n^b
        -{\tilde \alpha}_n^I {\tilde S}^b_{-n} )
\right)\,,
\nonumber\\
\\
\label{qt-}
 \sqrt{ 2 |p_-|}\,  {\tilde Q}_{\dot a}
&=& p_0^I \gamma^{I}_{\dot a b} {\tilde S}_0^b
  + |m| x_0^I\left(\gamma^I\Pi\right)_{\dot a b}{ S}_0^b
\nonumber\\
&+& \sum_{n=1}^\infty
\left(c_n  \gamma^{I}_{\dot a b}
       ({\tilde \alpha}_{-n}^I {\tilde S}_n^b
        +{\tilde \alpha}_n^I {\tilde S}^b_{-n} )-
 \frac{{\rm i} m}{2 \omega_n c_n }
 \left(\gamma^I\Pi\right)_{\dot a b}
       (\alpha_{-n}^I S_n^b - \alpha_n^I S^b_{-n})
\right)\,,
\nonumber\\
\end{eqnarray}
with $c_n = \sqrt{(\omega_n + n)/2\omega_n}$.
The combinations $Q^\pm={1\over \sqrt{2}}(Q\pm i\tilde Q)$
satisfy the anti-commutation relations
\bea
&&\left\{Q^+,  Q^-\right\} = 2
h+m(\gamma^{ij}\Pi)J^{ij}+m(\gamma^{i'j'}\Pi)J^{i'j'}\,,\nn\\
&&\left\{Q^+,    Q^+\right\} = \frac{1}{|p_-|} (N-\tilde N)\sim 0\, ,\nn\\
&& \left\{Q^-,    Q^-\right\} = \frac{1}{|p_-|}(N-\tilde N) \sim 0 \, ,
\label{dynsusyy}
\eea
where  $J^{IJ}$ is the generator
of angular momentum and $N,\tilde N$ are the left and right moving
number operators, defined by
\be
N= \sum_{n=1}^\infty\left( \frac{n}{\omega_n}\, \alpha_{-n}^I\alpha_{n}^I+ n\,
S^a_{-n} S^a_{n}\right) \, ,
\qquad
\tilde N= \sum_{n=1}^\infty \left(\frac{n}{\omega_n}\, \tilde
\alpha_{-n}^I\tilde \alpha_{n}^I +n\,\tilde S^a_{-n}\tilde S^a_n\right) \, .
\label{ndefs}
\ee
In the above and in what follows, a capital index $I,J,\dots$  labels
an $SO(8)$ vector, an unprimed lower case index  $i,j,\dots$ labels a
vector in one of the  $SO(4)$ subgroups, while a primed lower case
index  $i',j',\dots$  labels a vector in the other $SO(4)$ subgroup.
The symbol $\sim$ indicates that $N-\tilde N$ vanishes for physical
states, which satisfy the level-matching condition.

\subsection{Review of the $D$-instanton boundary state}

The $D$-instanton boundary state of \cite{gg} preserves eight
kinematical and eight dynamical supersymmetries and is given by
\be
\|{\bfz} \rangle\!\rangle = {\cal N}_{(0,0)}
\exp \left( \sum_{k=1}^{\infty}
\frac{1}{\omega_k} \alpha^I_{-k} \tilde\alpha^I_{-k}
- i {\eta}  S_{-k}M_k\tilde{S}_{-k}\right) \,
\|  {\bfz}\,\rangle\!\rangle_0\,,
\label{boundone}
\ee
where $\|  {\bfz}\,\rangle\!\rangle_0$ is the ground state of all
the oscillators of non-zero mode number. The coordinate ${\bfz}^I$
is the eigenvalue of the position operator,
\be
x_0^I = \frac{a^{\dagger\, I} -  a^I}{i\sqrt{2|m|}} \, ,
\label{harmosc}
\ee
constructed from the zero mode oscillators, $a^{\dagger\, I}$ and $ a^I$.
The parameter $\eta$ is equal to $\pm 1$, depending on whether the state
describes a $D$-instanton or an anti $D$-instanton.  From here on we
shall choose
$\eta=1$.  The
normalisation constant in (\ref{boundone}) is given by
\be\label{norma}
{\cal N}_{(0,0)} = (4\,\pi\, m)^2\,.
\ee
The matrix
\be\label{M}
M_k={1\over k}(\omega_k\bbbone- m \Pi)
\ee
satisfies
\be\label{orthogonal}
M_k M^T_{-k} = \bbbone \,.
\ee
The zero-mode part of the state is given by
\be
\| {\bfz} \rangle\!\rangle_0
= e^{-|m| {\bfz}^2 /2}  e^{i\sqrt{2|m|}{\bfz}\cdot {\bfa^\dagger}}
e^{\half {\bfa^\dagger}\cdot {\bfa^\dagger} } |0\rangle_D \,,
\label{boundzero}
\ee
where  $|0\rangle_D$ is the ground state of all the oscillators in the
basis in which $\theta_L$ and $\theta_R$ are annihilation modes.

The boundary state  was shown in \cite{gg} to satisfy the conditions
\be\label{fermions1}
\left(S^a_n + i \, M^{ab}_n\,
 \tilde{S}^b_{-n} \right)
\| {\bfz}\,\rangle\!\rangle = 0  \,.
\ee

It will be convenient to decompose the $SO(8)$ spinors  $S_n$ and
$\tilde{S}_n$ into spinors of definite $SO(4)$ chiralities by defining
\be\label{plusminus}
S^+_n = {1\over 2} (1+\Pi) S_n\,, \qquad
S^-_n = {1\over 2} (1-\Pi) S_n\,,
\ee
so that $\Pi S^\pm_n = \pm S^\pm _n$, and similarly for
$\tilde{S}_n$. Then (\ref{fermions1}) can be rewritten as
\be
\left(S^\pm_n + i \,M_n^{\pm} \tilde{S}^\pm_{-n} \right)
\| {\bfz}\rangle\!\rangle = 0 \,,
\ee
where
\be\label{rpmdef}
M_n^{\pm} = {\omega_n \mp \, m \over n}
= \sqrt{{\omega_n \mp \,m \over \omega_n \pm \, m}}\,.
\ee
Note also that
\be
{}_S\langle 0| S^{\pm}_{k} S^{\pm}_{-n}|0\rangle_S =\delta_{kn}\,,\qquad
{}_S\langle 0| S^{\pm}_{k} S^{\mp}_{-n}|0\rangle_S =0\,,
\ee
where $|0\rangle_S$ satisfies $S_k|0\rangle_S=\tilde S_k|0\rangle_S=0$ for $k> 0$.

The $n=0$ condition in (\ref{fermions1}) ensures that the state preserves half
the kinematical supersymmetries,
\be
q_L \|{\bfz}\,\rangle\!\rangle
=0 = q_R \| {\bfz}\,\rangle\!\rangle\, .
\label{kinsusy}
\ee
  Likewise, it preserves a
linear combination of dynamical supersymmetries,
\be
Q^+ \| {\bfz}\,\rangle\!\rangle = 0\, ,
\label{dynsusy}
\ee
where $\sqrt{2}Q^+ = Q + i \tilde Q$.   Applying the eight broken kinematical
supersymmetries,  $\bar q_L$ and $\bar q_R$,
and the eight broken dynamical supersymmetries,
 $\sqrt{2}Q^-=Q-i\tilde Q$,  to the boundary state generates sixteen fermionic moduli.

\subsection{Two states coupling to a $D$-instanton}

In this  subsection we will calculate the contribution to the mass
matrix of  two-string states coupling to the $D$-instanton.  To
leading order in the string coupling this process is determined by the
product of the one-point functions of closed strings coupling to a
disk world-sheet, as shown in figure 1.   It will be crucial to
include the sixteen fermionic moduli associated with open strings
coupling to the boundaries of the disks, which  were not discussed in
\cite{gg}.

\begin{center}
\FIGURE[ht]{
\epsfig{file=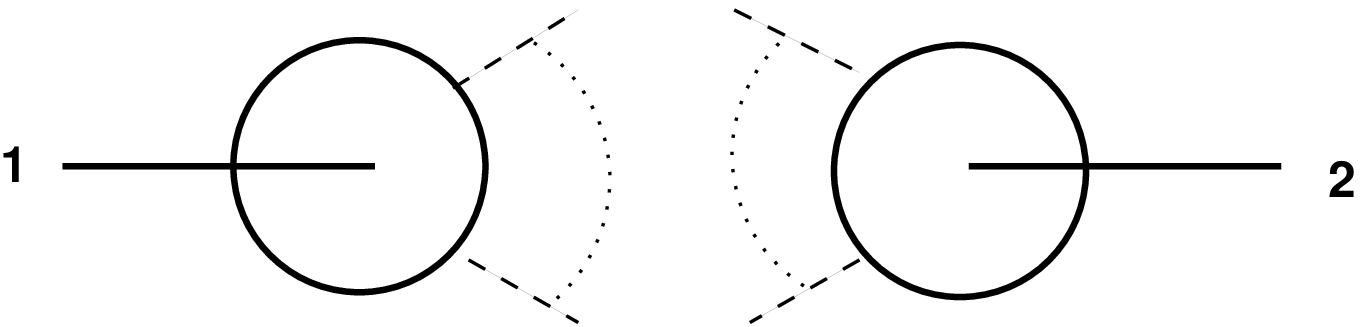,width=10.5cm, height=3cm}
\caption{Leading order contribution to the two-point function. Dashed lines
indicate the sixteen fermionic moduli, while solid lines indicate
external states. The complete process is a sum of such contributions
with the sixteen fermionic moduli distributed between the two disks in
all possible ways.}}
\end{center}

Each disk is defined in a separate Fock space labelled $1$ and
$2$, so that the $D$-instanton state in this space is given
by
\bea
\|\hat V^{(0)}_2\, , {\bfz}\rangle\!\rangle
&=&\|  {\bfz}\,\rangle\!\rangle_1 \otimes
\|  {\bfz}\,\rangle\!\rangle_2\, e^{i  x_0^+ (p_{1+} + p_{2+})}
e^{i x_0^- (p_{1-} + p_{2-})} \, ,
\label{dstate}
\eea
where the superscript
$(0)$ indicates that the fermionic moduli are not yet included.

Conservation of $p_-$ follows upon integrating over the modulus $x_0^-$
so that there is a factor of $\delta(p_{1-} + p_{2-})$ in the
two-point function.
We will take $p_{1-} \equiv p_- = -p_{2-} >0$, which
means that $m\equiv \mu\alpha' p_- >0$ on disk $1$ and $m <0$ on disk $2$.
After integration over the light-cone time modulus,
$x_0^+$, the process will preserve the $p_+$ component
of momentum, which means that there is a factor of $\delta(h_i + h_f)$,
where $h_i$, $h_f$ are the light-cone
hamiltonians for the incoming and outgoing states, respectively.

The dependence of the two-boundary state (\ref{dstate}) on the
transverse position modulus, ${\bfz}$, is given by the product of a
factor (\ref{boundzero}) for each disk.  Therefore,  making use of the
gaussian integral
\be
\int d^8 z
e^{-|m|{\bfz}^2} e^{-i\sqrt{2|m|}({\bfa^\dagger_ 1 + {\bfa^\dagger_ 2})
\cdot {\bfz}}} e^{({\bfa^\dagger_ 1}^2 + {\bfa^\dagger_ 2}^2)/2}
|0\rangle_1\otimes | 0\rangle_2
= \pi^4 m^{-4}
 e^{-{\bfa^\dagger_ 1}\cdot {\bfa^\dagger_2}}|0\rangle_1\otimes
 |0\rangle_2\,,
\ee
the two-particle vertex integrated over the bosonic moduli
(still ignoring fermionic moduli) has the form
\bea
&&\|\hat V^{(0)}_2\rangle\!\rangle \equiv
\int d^8 z \,  \|\hat V^{(0)}_2\, , {\bfz}\rangle\!\rangle\nn\\
&&= (2\pi)^8
\exp\left( \sum_{k=1}^{\infty}
\frac{1}{\omega_k} \alpha^{(1)I}_{-k} \tilde\alpha^{(1) I}_{-k}
- i  S^{(1)}_{-k}M_k\tilde{S}^{(1)}_{-k}\right. \nn\\
&& \left. \qquad \qquad\qquad +
\frac{1}{\omega_k} \alpha^{(2)I}_{-k} \tilde\alpha^{(2)I}_{-k}
- i  S^{(2)}_{-k}M_k\tilde{S}^{(2)}_{-k}\right)
 e^{-{\bfa^\dagger_ 1}\cdot {\bfa^\dagger_2}} |0
\rangle_1 \otimes |0 \rangle_2\,.
\label{zinnt}
\eea
Note, in particular, that the factor of $m^4$ in the normalization of
the boundary state cancels after integration over ${\bfz}$.

We will now consider the effect of applying the broken supersymmetries
acting on the $D$-instanton in order to determine the dependence on the
supermoduli.  The broken kinematic supersymmetries multiply the vertex by
the factor
\be
(\bar\epsilon_R(\bar q_{1R}  +\bar q_{2R}))^4\quad
(\bar\epsilon_L (\bar q_{1L}  +\bar q_{2L}))^4\,.
\label{brokkinsusy}
\ee
while the broken dynamical supersymmetries give the factor
\be
(\eta (  Q^-_1 +   Q^-_2))^8\, .
\label{brokdynsusy}
\ee
The spinors $\bar\epsilon_R^a$ and $\bar\epsilon_L^a$ are parameters of
the kinematical supersymmetries with opposite  $SO(4)\times SO(4)$
chiralities, so the index
$a$ has effectively four components for each chirality (which will be
labelled by $a_L$ and $a_R$).  The dynamical supersymmetry parameter
$\eta^{\dot a}$ has eight components.

Applying the broken kinematic supersymmetries (\ref{brokkinsusy})
to the original $D$-instanton state and integrating over
$\bar\epsilon_R$ and $\bar \epsilon_L$  gives the state
\bea
\|\hat V_2 \rangle\!\rangle
 &=& \epsilon_{a_{L_1}a_{L_2}a_{L_3} a_{L_4}}\,
\epsilon_{b_{R_1}b_{R_2}b_{R_3} b_{R_4}} \,
\prod_{r=1}^4 (\bar\theta_{2L}+\bar\theta_{1L})^{a_{L_r}}
(\bar\theta_{2R}+\bar\theta_{1R})^{b_{R_r}} \,
\|\hat V^{(0)}_2 \rangle\!\rangle \, .
\label{v2def}
\eea
The  products involving
$\bar\theta_L$'s and $\bar\theta_R$'s  can be interpreted as follows.
The boundary state  is located at a particular value of
${\bfz}, \epsilon_L, \epsilon_R$ in superspace.
The two disks are therefore associated with factors
$$\delta^4(\bar\theta_{1L}+\bar\epsilon_L) \delta^4(\bar\theta_{2L} +
\bar\epsilon_L)\,  $$
and a similar one involving $\bar\theta_R$'s.
Integrating over $\bar\epsilon_L$ and $\bar\epsilon_R$ gives the factor
$$\delta^4(\bar\theta_{2L}+ \bar\theta_{1L})\delta^4(\bar\theta_{2R}+
\bar\theta_{1R})
=  \epsilon_{a_{L_1}a_{L_2}a_{L_3} a_{L_4}}\,
\epsilon_{b_{R_1}b_{R_2}b_{R_3} b_{R_4}} \, \prod_{r=1}^4 (\bar\theta_{2L}+\bar\theta_{1L})^{a_{L_r}}
  (\bar\theta_{2R}+\bar\theta_{1R})^{b_{R_r}} \,.$$

The vertex $\|\hat V_2\rangle\!\rangle$   satisfies the conditions
\bea
({\bfa_1 + \bfa^\dagger_2})\|\hat V_2\rangle\!\rangle
 =&0&= ( {\bfa^\dagger_ 1 + \bfa_2}) \|\hat V_2\rangle\!\rangle
\nn\\
 (\bar\theta_{2L}+\bar\theta_{1L})\|\hat V_2\rangle\!\rangle
 =&0 &=
(\bar\theta_{2R} + \bar\theta_{1R})\|\hat V_2\rangle\!\rangle \, .
\label{condvv}
\eea

The remaining supermoduli are the $\eta^{\dot a}$ associated with the
broken dynamical supersymmetries, $Q^-$.  Applying the broken
dynamical supersymmetries to the state $\|\hat V_2\rangle\!\rangle $
produces an additional prefactor, resulting in the complete boundary
state,
\be
\| V_2\rangle\!\rangle =
\int d^8\eta\,  (\eta(Q^-_1 + Q^-_2))^8\|\hat V_2\rangle\!\rangle
 \, ,
\label{prefbrok}
\ee
which couples to any pair of physical closed-string states subject to
them preserving $p_+$ and $p_-$.

We now need to show that the unbroken supersymmetries,
\be
\epsilon_R^{a_R} (q_{1R} +q_{2R})^{a_R}\, \qquad
\epsilon_L^{a_L}(q_{1L}  +q_{2L})^{a_L}\,,
\qquad  \tilde\eta^{\dot a} (Q_1^+ -Q_2^+)^{\dot a}\, ,
\label{unbroks}
\ee
annihilate the vertex, $\| V_2\rangle\!\rangle$,
so the state preserves half the
supersymmetries.  The vertex $\|\hat V^{(0)}_2\rangle\!\rangle$ is automatically
annihilated by these supersymmetries, so the issue is whether they continue to
do so in the presence of the prefactors in (\ref{v2def}) and (\ref{prefbrok}).
So we need to show that the commutators of the unbroken supersymmetries with these
prefactors all vanish.   Although the conserved
kinematical supersymmetry,
$\epsilon_R (q_{1R}+q_{2R})$,
does not commute with the prefactor
$\left(\eta(Q_1^-+Q_2^-)\right)^8$, the commutator is proportional to
$(\bfa_1+\bfa_2^\dagger)$, which vanishes when acting on the vertex,
$\|\hat V_2\rangle\!\rangle$ (similarly, the commutator of the conserved
kinematical supersymmetry, $\epsilon_L (q_{1L}+q_{2L})$,  is
proportional to $(\bfa_2+\bfa_1^\dagger)$, which
also vanishes on the vertex).  The conserved dynamical supersymmetry,
$\eta\, (Q_1^+ - Q_2^+)$,
also does not commute with the prefactor in (\ref{prefbrok}), but the commutator
is proportional to the sum, $p_{1 +} + p_{2 +}$, which vanishes since
the light-cone energy is conserved in the on-shell two-point function.
Finally, the conserved dynamical supersymmetry does not commute
with the prefactor $\prod_{r=1}^4
(\bar\theta_{2L}+\bar\theta_{1L})^{a_{L_r}}
(\bar\theta_{2R}+\bar\theta_{1R})^{b_{R_r}}$ in  the vertex $\|\hat
V_2\rangle\!\rangle$, but its commutator is proportional to $({\bfa_1
+ \bfa^\dagger_2})$ or $({\bfa^\dagger_ 1 + \bfa_2})$, each of which vanishes when
acting on the vertex.

Elements of the mass matrix for external states $\chi_1$ and $\chi_2$
are proportional to  matrix elements of the form
\be
e^{2i\pi\tau}\, g_s^{7/2}\,
{}_1\langle \chi_1|\otimes {}_2\langle \chi_2\|V_2
\rangle\!\rangle \, .
\label{massels}
\ee
The analysis of the integral over the eight components of $\eta$ is
very different in the zero mode sector (the supergravity sector) from
the non zero-mode sector.  We will therefore first analyse the mass
matrix for the supergravity sector before considering more general
stringy effects.

\subsubsection{Decoupling of supergravity modes}

The piece of $\hQ$ that depends on zero modes is given by (see appendix)
\bea
\sqrt{2 |p_-|} Q^-_{0\dot a}&=&(p^I_0\gamma^I-i|m|x_0^I\gamma^I\Pi)_{\dot a b}
(S_0-i{\tilde S}_0)^b\nn\\
&=&
\sqrt{2}(p^I_0\gamma^I-i|m|x_0^I\gamma^I\Pi)_{\dot a b}
({\bar \theta}_R+{\bar \theta}_L)^b\nn\\
&=& 2\sqrt{|m|}\left({\bfa}\cdot\gamma{\bar
  \theta}_R+{\bfa^\dagger}\cdot\gamma{\bar \theta}_L\right)_{\dot a}\,,
\label{qmodess}
\eea
where we have used $\Pi{\bar \theta}_R=+{\bar \theta}_R$ and $\Pi
{\bar \theta}_L=-{\bar \theta}_L$.  We also note that the part of the
conserved dynamical supersymmetries, $Q^+$, that depends on zero modes
is given by
\be
\sqrt{2|p_-|} Q^+_{0{\dot a}}=2\sqrt{|m|}({\bfa} \cdot \gamma
\theta_L+{\bfa^\dagger}\cdot\gamma\theta_R)_{\dot a}\,.
\ee

The commutation relations between the $a$'s and $\hQ$ are given by
\be
[Q^-,{ a}^I]={\sqrt{2\mu}}\gamma^I {\bar \theta}_L,
\quad [Q^-,{a}^I]={\sqrt{2\mu}}\gamma^I {\bar \theta}_R\,, \label{bosQ}
\ee
and
\be
\{ Q^-,{\bar \theta}_L\}=0,\quad\{Q^-,{\bar \theta}_R\}=0\,.
\ee

The expressions for the zero-mode parts of the
broken dynamical supersymmetries that enter in (\ref{prefbrok})
are{\footnote{Note that according to our conventions an incoming state
    has positive momentum while an outgoing state has negative momentum.}}
\be
\hQ_1 = \sqrt{2\mu}\gamma\cdot \left({\bfa_1} \bar\theta_{1R} +
{\bfa^\dagger_ 1} \bar \theta_{1L}\right)\, , \qquad
\hQ_2  =  -\sqrt{2\mu}\gamma\cdot\left({\bfa^\dagger_ 2}
\bar\theta_{2R} + {\bfa_2} \bar \theta_{2L}\right)\,
\label{broksyr}
\ee

According to the above, the matrix elements of the mass matrix
between states
$\langle\chi_1|$ and $\langle\chi_2|$ have the form
\be
 {_{{}_1}}\langle \chi_1|\otimes {_{{}_2}}\langle\chi_2
\| V_2\rangle\!\rangle  =
\int d^8 \eta\, \, {_{{}_1}}\langle \chi_1|\otimes {_{{}_2}}\langle\chi_2|
\prod_{r=1}^8 (\eta^{\dot a_r}\, (Q^-_1  + Q^-_2)^{\dot a_r})
 \|\hat V_2\rangle\!\rangle\, .
 \label{matrixdef}
\ee

The last factor in $\prod \eta\,(Q^-_1+Q^-_2)$ acts on the boundary
state. In this factor one can substitute ${\bfa^\dagger_ 2}
=-{\bfa_1}$ and ${\bfa_1} = -\ {\bfa^\dagger_ 2}$ (using
(\ref{condvv})). This factor is then of the form
\be
\gamma\cdot \left(-{\bfa^\dagger_ 2}(\bar\theta_{1R}+\bar\theta_{2R})
+  {\bfa^\dagger_ 1} (\bar\theta_{1L}+ \bar\theta_{2L})\right)\, ,
\label{lastfac}
\ee
which vanishes when multiplying the
prefactor  $\prod_{r=1}^4(\bar\theta_{1L}+\bar\theta_{2L})^{{a_L}_r}
(\bar\theta_{1R}+\bar\theta_{2R})^{{b_R}_r}$ in $\|\hat V_2\rangle\!\rangle$.
We therefore see that the matrix elements involving two supergravity states
of arbitrary  excitation number vanish, as expected.

When one of the states has no stringy excitations and the other does the
matrix element also vanishes.  This follows from the fact that the two external
states have to have same value of $p_+$ (or level number) after integration
over the $x_0^+$ modulus of the $D$-instanton.

When both of the states have stringy excitations the matrix element does not
vanish.

\section{Matrix elements between  two-impurity single-string states}

We now turn to consider the matrix elements of states with non-zero
mode number excitations, which are generically non-vanishing.  One
general feature of these matrix elements follows from the fact that,
apart from the prefactor associated with the broken supersymmetries,
the boundary state is simply an exponential of a scalar quadratic form
in the excited oscillators.  In the absence of the prefactor the
boundary state would factorise into the product of two scalar
operators and therefore would only couple to states in which the
impurities are combined into $SO(4)  \times SO(4)$ singlets.  Only the
prefactor $ (\eta\, (Q^-_1 + Q^-_2))^8$ couples the $SO(4)\times
SO(4)$ spin between the two disks. In the first instance we will restrict
ourselves to states with only two impurities.

The complete  list of  two-impurity states based on the BMN vacuum at
a given mass level is given by
\ba
{NS-NS}~\qquad&:&
 \alpha^{(I}_{-n}\tilde \alpha^{J)}_{-n}|0\rangle_h\,, \qquad
 \alpha^{[I}_{-n}\tilde \alpha^{J]}_{-n}|0\rangle_h\, ,\qquad
\alpha^I_{-n}\tilde \alpha^I_{-n}|0\rangle_h\,,
\nn \\
{RR}~\qquad&:&  S^a_{-n}\tilde S^a_{-n}|0\rangle_h\, ,\qquad
S^a_{-n}\gamma^{IJ}_{ab}\tilde S^b_{-n}|0\rangle_h\, , \quad
S^a_{-n}\gamma^{IJKL}_{ab}\tilde S^b_{-n}|0\rangle_h\,,\nn\\
{NS-R}~\qquad&:& \alpha^I_{-n}\tilde S^a_{-n}|0\rangle_h\,, \nn\\
{R-NS}~\qquad&:& S^a_{-n}\tilde \alpha^I_{-n}|0\rangle_h\,,
\label{statelist}
\ea
where $NS$ and $R$ indicate the Neveu--Schwarz and Ramond sectors,
respectively.  The masses of each of these states is $2\omega_n$ in
the free string theory, while our aim is to evaluate the one-instanton
mass matrix that corrects the masses of such states.  Although the
list (\ref{statelist}) is labelled with SO(8) transverse vector
indices for economy of presentation, the boundary state only respects
the $SO(4) \times SO(4)$ subgroup.  We will first  consider states in
the $NS-NS$ sector that have two bosonic impurities.

\subsection{Two bosonic impurities}

In this case we will take the external states
to be stringy excitations of the BMN ground state.  In this case
the bra vectors in (\ref{matrixdef}) are given by
\be
 {_{{}_1}}\langle \chi_1|\otimes {_{{}_2}}\langle\chi_2| =
 \frac{1}{\omega_p \omega_q}\, t_{IJ}^{(1)}\, t_{KL}^{(2)}
 { }_h\langle 0|\alpha_p^{(1) I}\, \tilde \alpha_p^{(1) J}
 \otimes  {}_h\langle 0|\alpha_q^{(2)K}\, \tilde \alpha_q^{(2)L}  \, .
\label{bostates}
\ee
The normalisation has been chosen so that each external state has unit
norm if the wave functions satisfy $t_{IJ}^{(1)}\, t^{(1)}_{IJ} =1
=t_{IJ}^{(2)}\, t^{(2)}_{IJ}$.

We now  proceed to evaluate the matrix element (\ref{matrixdef}).
The non-zero modes enter into $Q^-$ in the following manner
\bea
\sqrt{2 |p_-|}{Q^-_{\dot
  a}}_{n\neq 0}&=&\sum_{n=1}^{\infty}(\gamma^I N_+)_{\dot
  a b}(\alpha^I_n S^b_{-n}-i\,{\tilde \alpha}^I_n{\tilde S}^b_{-n})\nn\\
&+&
(\gamma^I N_-)_{\dot
  a b}(\alpha^I_{-n} S^b_{n}-i\,{\tilde \alpha}^I_{-n}{\tilde
  S}^b_{n})\,,
\label{muQexact}
\eea
where we have defined
\be
  (N_\pm)_{ab}= (c_n\bbbone\pm {m\over 2\omega_n c_n}\Pi)_{ab}\,.
\ee
The matrices $N_+$ and  $N_-$ satisfy
\be
N_+^2={n\over \omega_n}M_{-n}\,, \qquad N_+ N_-={n\over \omega_n}\,, \qquad
N_-^2={n\over \omega_n} M_n\,.
\ee
We will make use of the following commutation relations that are valid
for $p_->0$,
\ba
\sqrt{2|p_-|}\left[\hQ_{\dot a},\alpha^J_{-p}\right]
&=&\omega_p\gamma^I N_{+{\dot a b}} S^b_{-p} \label{com1exact}\\
\sqrt{2|p_-|}\left[\hQ_{\dot a},{\tilde \alpha}^J_{-p}\right] &=& -i\omega_p
\gamma^I N_{+{\dot
  a b}}{\tilde S}^b_{-p} \label{com2exact}\\
\sqrt{2|p_-|}\left\{\hQ_{\dot a},S^b_{-p}\right\}&=&\gamma^I N_{-{\dot
  a b}}\alpha^I_{-p}\label{com3exact}\\
\sqrt{2|p_-|}\left\{\hQ_{\dot a},{\tilde S}^b_{-p}\right\}
&=&-i\gamma^I N_{-{\dot a b}}{\tilde \alpha}^I_{-p}\,.
\label{com4exact}
\ea
For $p_-<0$ the sign of $m$ changes and the matrices $N_+$ and $N_-$
are interchanged.

The vertex $\| V_2\rangle\!\rangle$ in (\ref{matrixdef}) contains
the prefactor
\be
(\eta (Q^-_1 + Q^-_2))^8 = \sum_{p=0}^8 C^8_p (\eta Q^-_1)^p
(\eta Q^-_2)^{8-p}\, ,
\label{dynprod}
\ee
where $C^8_p$ are binomial coefficients.  The different terms in the
sum in (\ref{dynprod}) generate couplings between external states in
different $SO(4) \times SO(4)$ representations when they act on the
boundary state $\| \hat V_2\rangle\!\rangle$.

With external states of the form (\ref{bostates}) an even number of
$Q^-$'s must be distributed between the disks so only the
terms with even $p$  contribute.  For each value of $p$ each disk
couples to an external string state that lies in symmetric,
antisymmetric or trace irreducible representations of $SO(4) \times
SO(4)$ which are shown in table 1.

{\begin{center}\begin{tabular}[ht]{|c|c|c|c|c|}
\hline
$p$ & Disk 1& Disk 2& $SO(4)\times SO(4)$ reps.\cr
\hline
0 & $(Q^-)^0$&  $(Q^-)^8$ & $(i,i) + (j',j')$\cr
2 & $(Q^-)^2$&  $(Q^-)^6$ &  $[i,j]\,, [i',j'] \,,  [i,j']$  \cr
4 & $ (Q^-)^4$&  $(Q^-)^4$ & $(i,j)_t\,, (i',j')_t\, , (i,i)\,,(j',j')
\, ,(i,j')  $  \cr
6 & $(Q^-)^6$&  $(Q^-)^2$  &  $[i,j]\,, [i',j'] \,,  [i,j']$  \cr
8 & $(Q^-)^8$&  $(Q^-)^0$  & $(i,i) + (j',j')$\cr
\hline
\end{tabular}\end{center}
{\bf Table 1: }{\small Distribution of the eight $Q^-$'s between the
    two disks leads to couplings between pairs of $SO(4)\times SO(4)$
    irreducible representations listed in the last column.   The
    symbol $[a,b]$ indicates an antisymmetric representation,
    $(a,b)_t$ indicates a symmetric traceless irreducible
    representation  while  $(a,a)$ indicates a singlet of either
    $SO(4)$.}  }

\subsubsection{Matrix elements of symmetric tensor and singlet states}

We will first consider the $p=4$ case, in which there is a pre-factor
of $Q^4$.  This contributes to the $SO(4)\times SO(4)$
representations, $(i,j)$, $(i',j')$, $(i,i)$, $(j',j')$ and $(i,j') +
(j',i)$.  In this case  we need to include the terms with binomial
coefficients $C^8_4$ in (\ref{dynprod}).

Since the external states are bra vectors containing two excited
annihilation oscillators, and since each factor of $(Q_1^- + Q_2^-)$
in the prefactor contains the sum of products of one creation mode and
one annihilation mode, we need only keep the bilinear terms in the
expansion of the exponential factor in the  boundary ket state.  The
matrix elements therefore have the form
\bea
 {_{{}_1}}\langle \chi_1|\otimes {_{{}_2}}\langle\chi_2|  &&
(\eta(Q_1^-+Q_2^-))^8|
\left( \sum_{k=1}^{\infty}\frac{1}{\omega_k} \alpha^{(1)I}_{-k}
\tilde\alpha^{(1)I}_{-k}
- i   S^{(1)}_{-k}M_k\tilde{S}^{(1)}_{-k}\right) \nn\\
&& \left( \sum_{l=1}^{\infty} \frac{1}{\omega_l}
\alpha^{(2)J}_{-l} \tilde\alpha^{(2)J}_{-l}
- i  S^{(2)}_{-l}M_l\tilde{S}^{(2)}_{-l}\right) |0
\rangle_1 \otimes |0 \rangle_2\,.
\label{finmat}
\eea

We will evaluate (\ref{finmat}) by commuting factors of
$\eta(Q_1^-+Q_2^-)$ to the right, noting that the commuted part
annihilates the ground-state ket vector.  We therefore pick up
factors of $[(\eta Q^-)^r, \alpha^I]$ or  $[(\eta Q^-)^r, S^a]$ for
various values of the integer $r$.  These factors are summarised in
table 2 (in which $N^I_\pm \equiv \gamma^I\, N_\pm$)\footnote{Recall
that the sign of $m$ in $Q_2^-$ is reversed relative to that in
$Q_1^-$.}.  Overall powers of $|p_-|$ are omitted in this list since
these  cancel with factors of $|p_-|$ coming from the kinematic
supersymmetries once both the disks are included.

\vskip 0.3cm
{\small
\hskip -1cm
{\begin{center}
\begin{tabular}
{|c|c|c|}\hline
& \rule[-6pt]{0pt}{18pt}\hspace*{-0.1cm}
$\alpha^I_{-n}$ & $S^a_{-n}$
 \\ \hline
$\rule[-6pt]{0pt}{18pt}\hspace*{-0.1cm}
\sqrt{2|p_-|}Q^{-\dot a_1}$ & $\omega_n(N^I_+)_{\dot a_1 c}S^c_{-n}$ &
  $(N^I_-)_{\dot a_1 a} \alpha^I_{-n}$ \\ \hline
$\rule[-6pt]{0pt}{18pt}\hspace*{-0.1cm}
\prod_{r=1}^2\sqrt{2|p_-|} Q^{-\dot a_r}$ & ${n}
  \gamma^{IJ}_{\dot a_1\dot a_2}
   \alpha^J_{-n}$ &  $\omega_n(N_-^I)_{\dot
  a_1 a}(N_+^I S_{-n})_{\dot a_2}$  \\ \hline
$\rule[-6pt]{0pt}{18pt}\hspace*{-0.1cm}
\prod_{r=1}^3 \sqrt{2|p_-|}Q^{-\dot a_r}$& ${n\omega_n}
\gamma^{IJ}_{\dot a_1 \dot a_2}(N_+^J S_{-n})_{\dot a_3}$ &
${n}(N_-^I)_{\dot a_1 a}\gamma^{IJ}_{\dot a_2 \dot
  a_3}\alpha^J_{-n}$  \\ \hline
$\rule[-6pt]{0pt}{18pt}\hspace*{-0.1cm}
\prod_{r=1}^4 \sqrt{2|p_-|}Q^{-\dot a_r}$& $n^2\gamma^{IJ}_{\dot
  a_1 \dot a_2}\gamma^{JK}_{\dot
  a_3 \dot a_4}\alpha_{-n}^K$& ${n
  \omega_n}(N_-^I)_{\dot a_1 a}\gamma^{IJ}_{\dot a_2 \dot
  a_3}(N_+^J S_{-n})_{\dot a_4}$ \\ \hline
$\rule[-6pt]{0pt}{18pt}\hspace*{-0.1cm}
\prod_{r=1}^5 \sqrt{2|p_-|}Q^{-\dot a_r}$& $\omega_n n^2
\gamma^{IJ}_{\dot a_1 \dot a_2}\gamma^{JK}_{\dot
  a_3 \dot a_4} (N_+^K S_{-n})_{\dot a_5}$ & $n^2(N_-^I)_{\dot a_1 a}
\gamma^{IJ}_{\dot a_2 \dot a_3}\gamma^{JK}_{\dot a_4 \dot
  a_5}\alpha_{-n}^K$\\ \hline
$\rule[-6pt]{0pt}{18pt}\hspace*{-0.1cm}
\prod_{r=1}^6 \sqrt{2|p_-|}Q^{-\dot a_r}$& $n^3\gamma^{IJ}_{\dot
  a_1 \dot a_2}\gamma^{JK}_{\dot
  a_3 \dot a_4}\gamma^{KL}_{\dot
  a_5 \dot a_6}\alpha_{-n}^L$& $\omega_n n^2(N_-^I)_{\dot a_1 a}
\gamma^{IJ}_{\dot a_2 \dot a_3}\gamma^{JK}_{\dot a_4 \dot
  a_5} (N_+^K S_{-n})_{\dot a_6}$ \\ \hline
$\rule[-6pt]{0pt}{18pt}\hspace*{-0.1cm}
\prod_{r=1}^7 \sqrt{2|p_-|}Q^{-\dot a_r}$&
$\omega_n n^3\gamma^{IJ}_{\dot
  a_1 \dot a_2}\gamma^{JK}_{\dot
  a_3 \dot a_4}\gamma^{KL}_{\dot
  a_5 \dot a_6}(N_+^L S_{-n})_{\dot a_7}$ & $n^3(N_-^I)_{\dot a_1 a}
\gamma^{IJ}_{\dot a_2 \dot a_3}\gamma^{JK}_{\dot a_4 \dot
  a_5}\gamma^{KL}_{\dot a_6 \dot
  a_7}\alpha^L_{-n}$ \\ \hline
$\rule[-6pt]{0pt}{18pt}\hspace*{-0.1cm}
\prod_{r=1}^8 \sqrt{2|p_-|}Q^{-\dot a_r}$&
$n^4\gamma^{IJ}_{\dot
  a_1 \dot a_2}\gamma^{JK}_{\dot
  a_3 \dot a_4}\gamma^{KL}_{\dot
  a_5 \dot a_6}\gamma^{LM}_{\dot
  a_7 \dot a_8}\alpha^M_{-n}$ & $\omega_n n^3(N_-^I)_{\dot a_1 a}
\gamma^{IJ}_{\dot a_2 \dot a_3}\gamma^{JK}_{\dot a_4 \dot
  a_5}\gamma^{KL}_{\dot a_6 \dot
  a_7} (N_+^L S_{-n})_{\dot a_8}$ \cr \hline
\end{tabular}
\end{center}}
\smallskip
\noindent
{\bf Table 2: }{\small Action of broken supersymmetries on the left-moving
(untilded) oscillators with
$p_->0$.  The result for the right-movers (tilded) oscillators is given by
inserting an extra factor of  $(-i)^r$.  When $p_-<0$
the matrices $N_-$ and $N_+$ are interchanged.  The matrices $N_\pm^I$
are defined by $N^I_\pm \equiv \gamma^I\, N_\pm$.}}

\vskip 0.3cm

Since the external states are made of bosonic oscillators only, we
will need to act with an even number of $Q^-$'s on each bosonic
oscillator and with an odd number on each fermionic oscillator.   In
other words, we need to evaluate an expression of the form,
\bea
&&{1\over \omega} [(\eta Q^-)^4,\alpha)]\tilde \alpha+{1\over
  \omega} \alpha
[(\eta Q^-)^4), \tilde \alpha] + 6{1\over \omega}[(\eta Q^-)^2,\alpha ]\
[(\eta Q^-)^2,\tilde \alpha] \nn\\
&&\qquad\qquad  +4 [(\eta Q^-)^3, S] \  [\eta Q^- ,\tilde S]+4 [\eta Q^-,
S]\  [(\eta Q^-)^3 ,\tilde S] \,,
\eea
where we have suppressed the index structure. The various numerical
factors in front of each term are the binomial coefficients $C^4_n$.

Using table 1 and the following identity
\be \label{id1}
N_- M_n N_-+{n\over \omega_n}=2 M_n\,,
\ee
we obtain a contribution to the first disk of the form
\be
|\!|{\rm Disk~1}\rangle\!\rangle\equiv \sum_{n=1}^\infty n \left[ (\eta\gamma^R
  \gamma^K\eta)\left(\eta\gamma^K ({\omega_n\over n}\bbbone-{m\over n}\Pi)
  \gamma^S\eta\right)\right]\alpha^{(1)(R}_{-n}
{\tilde \alpha}^{(1) S)}_{-n}|
0\,\rangle_1\,.
\label{4Q}
\ee
For the second disk, the result can be obtained from equation
(\ref{4Q}) by replacing $m$ by $-m$, giving
\be
|\!|{\rm Disk~2}\rangle\!\rangle\equiv  \sum_{n=1}^\infty n\left[ (\eta\gamma^P
\gamma^L\eta)\left(\eta\gamma^L ({\omega_n\over n}\bbbone+{m\over n}\Pi)
\gamma^Q\eta\right)\right]\alpha^{(2)(P}_{-n}{\tilde \alpha}^{(2)Q)}_{-n}
|0\,\rangle_2\,.
\ee
The mass matrix is obtained by evaluating
\be\label{answer}
\frac{1}{\omega_p \omega_q}t^{(1)}_{(IJ)}t^{(2)}_{(KL)}
\int d^8 \eta\,\langle 0|
\alpha^{(1)I}_{p}\tilde\alpha^{(1)J}_p
|\!|{\rm Disk~1}\rangle\!\rangle\times
\langle 0|\alpha^{(2)K}_q\tilde\alpha^{(2)L}_q|\!|{\rm Disk~2}\rangle\!\rangle\,,
\ee
This expression involves the Grassmann integral,
\be
\int d^8 \,\eta \eta \gamma^{IK}\eta \, \eta \gamma^K M_n^- \gamma^J \eta \,
\eta \gamma^{PL} \eta \, \eta \gamma^L M_n^+ \gamma^Q \eta\,.
\ee
Integration over the $\eta$'s can be efficiently
expressed in terms of the four-component chiral $SO(4)\times
SO(4)$ spinors
\be
\eta^\pm = \frac{1}{2} ({1\pm \Pi})\, \eta\, .
\label{etachir}
\ee
The integrals we will meet later are all of the form
\ba
\int d^4 \eta^{\pm} \eta^{\pm}\gamma^{ij}\eta^\pm\,
\eta^\pm\gamma^{kl}\eta^\pm&=&(-\delta^{ik}\delta^{jl}+\delta^{il}
\delta^{jk}\pm\epsilon^{ijkl})\equiv {\calT}^\pm_{ijkl}\,,\label{etaints1}\\
\int d^4 \eta^{\pm} \eta^{\pm}\gamma^{i'j'}\eta^\pm\,
\eta^\pm\gamma^{k'l'}\eta^\pm&=&(\delta^{i'k'}\delta^{j'l'}-\delta^{i'l'}
\delta^{j'k'}\pm\epsilon^{i'j'k'l'}) \equiv -{\calT}^\mp_{i'j'k'l'}\,,
\label{etaints2}\\
\int d^4 \eta^{\pm} \eta^{\pm}\gamma^{ij}\eta^\pm\,
\eta^\pm\gamma^{k'l'}\eta^\pm&=& 0\,,
\label{etaints3}
\ea
or can be converted to this form by making use of the Fierz transformation
\be
\eta^{\pm}_{\dot a}\eta^{\pm}_{\dot b}={1\over
  16}\left((\gamma^{kl}P^\pm)_{\dot a\dot b}\eta
\gamma^{kl}\eta+(\gamma^{k'l'}P^\pm)_{\dot a\dot
  b}\eta\gamma^{k'l'}\eta\right)\,,
\label{fierzeta}
\ee
where $P^\pm={1\over 2}(\bbbone\pm \Pi)$.

The tensors $\calT^\pm_{ijkl}$,
defined in (\ref{etaints1}),  have the property that
they are left invariant under the interchange of the first and last
pair of indices and they satisfy the (anti)self-duality condition
\be
\epsilon^{j_1 j_2 p_3 p_4}\, {\calT}^\pm_{j_1 j_2 j_3 j_4}
=\pm 2 {\calT}^\pm_{p_3 p_4 j_3 j_4}\,.
\ee
We also note the property
\be
\calT^+_{ispj}\calT^-_{rijq}=2(\delta^{rq}\delta^{ps}+\delta^{sq}\delta^{rp})-
\delta^{pq}\delta^{rs}\, , \label{prodT}
\ee
which is invariant
under $p\leftrightarrow q, r\leftrightarrow s$.

\smallskip
\smallskip

\noindent {\bf   Mass matrix elements with $(I,J)=(i,j)$}
\smallskip

Up to now the indices $I,J,K,L$ have labelled $SO(8)$ vectors which
can take values in either of the $SO(4)$ factors in the $SO(4)\times
SO(4)$ subgroup.  At this point we will specialise to the
representation in which the vector indices $(I,J)$ are in one of the
$SO(4)$ subgroups of $SO(4) \times SO(4)$, so that $(I,J) \to (i,j)$.
In this case the integration over the fermionic moduli, $\eta$, in
({\ref{answer}) involves evaluation of the Grassmann integral
\be
\int d^8 \,\eta \eta \gamma^{iK}\eta \, \eta \gamma^K M_n^- \gamma^j \eta \,
\eta \gamma^{PL} \eta \, \eta \gamma^L M_n^+ \gamma^Q \eta\,,
\ee
where
\be
M_n^\pm={\omega_n \over n}1\pm{m\over n}\Pi\, .
\label{mpmdef}
\ee

We will only consider the leading power of $m$ in the $m\to \infty$
limit, which is relevant to the comparison with the gauge theory.  The
only non-zero matrix elements arise when the indices $(P,Q)$ are in
the first $SO(4)$ factor, so that $(P,Q) = (p,q)$.  In this case
(\ref{fierzeta}) and (\ref{etaints1}) lead to (in the limit $m\to
\infty$)
\bea
&&\frac{n^2}{m^2} \int d^8 \,\eta\,\eta \gamma^{iK}\eta \, \eta
\gamma^K M_n^- \gamma^j \eta \,
\eta \gamma^{pL} \eta \, \eta \gamma^L M_n^+ \gamma^q \eta  \nn\\ &&  \to
({\calT}^+_{ikkj}{\calT}^-_{pllq})+(9{\calT}^-_{iklq}{\calT}^+_{kjpl})
 = 9(\delta^{pq}\delta^{ij})+18(\delta^{pi}\delta^{qj}+\delta^{pj}
\delta^{iq}-{1\over 2} \delta^{pq}\delta^{ij})\,.
\label{specel}
\eea
This expression   couples the symmetric traceless
wave functions, $t^{(1)}_{(ij)_t}$ and $t^{(2)}_{(pq)_t}$
as well as the $SO(4)$ singlets $t^{(1)}_{(ii)}$ and $t^{(2)}_{(pp)}$.
Matrix elements with $(I,J)= (i,j)$ and  $(P,Q) = (p',q')$ vanish.
Of course, there is an expression similar to (\ref{specel}) with primed indices
replacing the unprimed ones.

To summarise, the matrix elements between states with two bosonic
impurities with symmetrised indices in the same $SO(4)$ factor are
proportional to
\be
e^{2\pi i \tau}\, g_s^{7/2}\, m^4\, t^{(1)}_{ij}\, t^{(2)}_{pq}\,
(\delta^{pi}\delta^{qj}+\delta^{pj}\delta^{iq})\, .
\label{summsymm}
\ee

\noindent {\bf Mass matrix elements with  $(I,J) = (i,j')$}

\smallskip

The disk with four  $Q^-$'s   attached couples to the
symmetric combination   $(I,J)\rightarrow (i,j')\oplus (j',i)$.
Explicitly, the $\eta$-dependence for disk 1 appears in the combination
\begin{equation}
{1\over 2}\eta\gamma^{iK}\eta\eta\gamma^K(1-\Pi)\gamma^{j'}
 \eta=\eta^+ \gamma^{ik}\eta^+ \eta^+ \gamma^{kj'}\eta^- +
\eta^-\gamma^{ik}\eta^-\eta^+\gamma^{kj'}\eta^-+2\eta^+
\gamma^{ik'}\eta^- \eta^-\gamma^{k'j'} \eta^-\,,
\label{diskone}
\end{equation}
which only has odd powers of $\eta^+$ and $\eta^-$. The
analogous factor for  disk 2 is
\begin{eqnarray}
 {1\over 2}\eta\gamma^{PL}\eta\eta\gamma^L(1+\Pi)\gamma^Q
& =&  \eta^+\gamma^{pl}\eta^+ \eta^+\gamma^{lq'}\eta^-+\eta^-
\gamma^{pl}\eta^-\eta^-\gamma^{lq'}\eta^+
  \nonumber \\
&+& 2 \eta^-\gamma^{pl'}\eta^+ \eta^+\gamma^{l'q'}\eta^+ + 2 \eta^+
\gamma^{p'l} \eta^- \eta^-
  \gamma^{lq}\eta^-\nn\\
  &+& \eta^+ \gamma^{p'l'} \eta^+ \eta^+ \gamma^{l'q} \eta^- + \eta^-
\gamma^{p'l'} \eta^-
  \eta^+\gamma^{l'q} \eta^-\,.
\label{disktwo}
\end{eqnarray}
Multiplying the expressions (\ref{diskone}) and (\ref{disktwo}) gives three
types of terms,
\begin{eqnarray}
&&\int  d^4 \eta^+ d^4\eta^- \left[ ( \eta^+\gamma^{ik} \eta^+
\eta^-\gamma^{kj'}\eta^+
\eta^+\gamma^{l'q}\eta^- \eta^- \gamma^{p'l'}\eta^-) \right. \nn\\
&&\left.+(\eta^-\gamma^{ik}\eta^- \eta^+\gamma^{p'l'}\eta^+
\eta^+\gamma^{kj'}\eta^- \eta^+
\gamma^{l'q}\eta^-) \right]
 = -{1\over 8} (\calT^+_{ikkq}\calT^-_{p'l'l'j'}+\calT^-_{ikkq}
\calT^+_{p'l'l'j'})\, ,
\end{eqnarray}
\be
2\int d^4 \eta^+ d^4 \eta^- \eta^-\gamma^{ik}\eta^- \eta^+
\gamma^{l'q'}\eta^+ \eta^-\gamma^{kj'} \eta^- \eta^-\gamma^{pl'}
\eta^+=-{1\over 4}\calT^-_{ikkp}\calT^+_{q'l'l'j'}\,,
\ee
\be
2\int d^4 \eta^+ d^4 \eta^- \eta^-\gamma^{k'j'}\eta^- \eta^+
\gamma^{pl}\eta^+ \eta^-
\gamma^{ik'} \eta^- \eta^+\gamma^{lq'} \eta^-=-{1\over 4}
\calT^-_{j'k'k'q'}\calT^+_{plli}\,.
\ee
It is straightforward to show from these expressions that
the matrix elements of this type are of the form
\be
e^{2\pi i \tau}\, g_s^{7/2}\, m^4\,
t^{(1)}_{ij'}\, t^{(2)}_{p'q} \left(\delta_{iq}\delta_{p'j'}
  +\delta_{q'j'}\delta_{pi}\right)\,.
\label{mixmat}
\ee

It is of interest to compare the above matrix elements with those
deduced from $\calN=4$ Yang--Mills theory in the BMN limit.  The gauge
theory parameters relevant to this limit are expressed in terms of
those of string theory by the relations \cite{bmn,kpss+7}
\be
\frac{J^2}{N} \equiv g_2 = 4\pi g_s m^2\, , \qquad g^2_{_{YM}}\frac
{N}{J^2} \equiv \lambda'=\frac{1}{m^2},
\label{gaugedef}
\ee
and the light-cone string theory hamiltonian is expressed in terms of
$J$ and $\Delta$ by $H^{(2)}/\mu = \Delta-J$.  The presence of the
$D$-instanton contribution to the two-particle hamiltonian in string
theory therefore implies that there should be a corresponding
contribution to the two-point function of the corresponding BMN
operators in $\calN=4$ Yang--Mills gauge theory of the form
$$
e^{i \theta -8\pi^2/g_{_{YM}}^2}  \,  g_2^{7/2} \, \lambda'^2\,.
$$
We will see in \cite{gks} that this dependence on the
coupling constant does indeed emerge from the gauge theory although it
proves very difficult to evaluate the instanton contribution in detail
in the two-impurity case\footnote{It turns out that in the gauge
theory, the four-impurity case is under better control.}.

\subsubsection{Matrix elements of $SO(8)$ antisymmetric tensors in
the $NS-NS$ sector}

The two-string state describing strings in the  antisymmetric $NS-NS$
representation is
\be
\frac{1}{\omega_p \omega_q}\, t_{[IJ]}^{(1)}\, t_{[KL]}^{(2)}\,
 { }_h\langle 0|\alpha_p^{(1) I}\, \tilde \alpha_p^{(1) J}
 \otimes  {}_h\langle 0|\alpha_q^{(2)K}\, \tilde \alpha_q^{(2)L}.
\label{antisymm}
\ee
In this case the boundary state contribution comes from terms in
(\ref{dynprod}) with $p=2$ and $p=6$, where two $Q^-$'s are
distributed on one disk and six  on the other.  Specialising again to
the case in which the vector indices in (\ref{antisymm}) lie in one of
the $SO(4)$ subgroups of $SO(8)$ the matrix element turns out to be
proportional to $(\calT^+_{ijkl} + \calT^-_{ijkl})$.  To leading order
in $m$ it is proportional to
\be
e^{2\pi i \tau}\,g_s^{7/2}\, m^4\,  t^{(1)}_{ij}t^{(2)} _{kl}
\, (\delta_{ik}\delta_{jl} - \delta_{jk}\delta_{il}) \, ,
\label{asres2}
\ee
where $t^{(1)}$ and $t^{(2)}$ are the antisymmetric tensor wave
functions of the two states.  The result (\ref{asres2}) has the same
dependence on the parameters as in the symmetric case considered
earlier.

A similar result follows when the external states have vector indices
in the other $SO(4)$.

It is also easy to see that there is no mixing of the $[i,j]$ states
with the $[i',j']$ states.  Furthermore a state with one index in each
of the  $SO(4)$ factors, $[i,j']$, only mixes with a similar state,
again resulting in a dependence on the parameters of the form $e^{2\pi
i \tau}\,g_s^{7/2}\, m^4$.

\subsubsection{ Matrix elements of $SO(8)$ singlets in the $NS-NS$ sector}

As remarked earlier, the singlet $SO(8)$ representation is the direct
sum of  $SO(4)$ singlets in $SO(4)\times SO(4)$.  This is denoted by
$(ii) + (j'j')$ in table 1.  In this case the  wave functions are
$t^{(1)}_{II}$ and $t^{(2)}_{II}$ and the result turns out to be
proportional to $g_s^{7/2}m^2$.  This is suppressed by a power of
$m^{-2}$ relative to the matrix elements of $(ii) - (j'j')$, which
accounts for the earlier observation that this matrix element vanishes
in the $m\to \infty$ limit.  In this case the gauge theory result
should be proportional to $g_{YM}^6$ rather than $g_{YM}^4$.

\subsection{Matrix elements of excited $RR$ states}

We will only briefly consider matrix elements of pairs of states with
two $R$ impurities. For simplicity let us consider the $SO(4)\times
SO(4)$ $RR$ singlet that comes from the $SO(8)$ $RR$ four-form. In
this case the leading contribution arises when each disk has four
$Q^-$'s. The dominant contribution in the large $m$ limit arises
when two $Q^-$'s act on each $S$ and $\tilde S$ on each disk. Consider
the case in which the external state on disk 1 (with $p_- >0$) is of
the form $S^+\tilde S^+$.  This disk contributes at order $m^3$.  Now
consider the case that the external state on the disk 2 (with $p_-<0$
so that $m < 0$)  is of the form $S^+\tilde S^+$.  This is suppressed
by $1/m^2$ relative to the first disk and the combined power of $m$ is
$m^4$, as in the $NS-NS$ sector.  Note that this is an example of a
two-point function which also gets perturbative contributions, as do
the $NS-NS$ two-point functions considered earlier.  However, if the
external state is of the form $S^-\tilde S^-$ the result is
proportional to $m^3$ on each disk and the net power of $m$ is $m^6$.
In this example there are no perturbative contributions, which follows
from the fact that  $\langle S^+ S^- \rangle =0$.

\subsubsection{Other matrix elements of two-impurity states}

The last example illustrates a general feature of states involving
fermionic excitations, namely, that the two-point functions of states
with a large number of fermionic excitations can have a high positive
power of $m$. On the gauge theory side, this corresponds to a large
negative power of $\lambda'$.  Of course, since the instanton
contributions under consideration have a prefactor of
$e^{-8\pi^2/g_{_{YM}}^2} = e^{-8\pi^2 N /\lambda' J^2}$ the
$\lambda'\to 0$ limit is not divergent.  We will see in \cite{gks}
that this qualitatively matches the behaviour of the analogous
Yang--Mills instanton contributions to the gauge theory in the BMN
limit.

An important general observation is that the $D$-instanton induces
mixing between $NS-NS$ and $R-R$ states which does not occur at any
order in string perturbation theory. This is easily seen from the fact
that the $D$-instanton boundary state is a source of $R-R$
charge. This mixing is discussed in more detail in \cite{gks2}.

Fermionic states can be analysed in a similar
manner.  They require an odd number of $Q^-$'s on each disk.

\section{Matrix elements between states with four bosonic impurities}
In this section we consider external states made from four bosonic
oscillators. It will turn out that the comparison of matrix elements
for these states with the anomalous dimensions of corresponding
operators in the gauge theory is under better control
than in the case of two impurity operators.  We will only consider the
case in which all the vector indices on the bosonic oscillators are in
one of the $SO(4)$ factors.  In this case the two-particle state has
the form
\bea
&& {_{{}_1}}\langle \chi_1|\otimes {_{{}_2}}\langle\chi_2| =
\frac{1}{\omega^2_r \omega^2_s}\, t_{j_1j_2j_3j_4}^{(1)}\,
t_{p_1p_2p_3p_4}^{(2)}\nn\\
 && { }_h\langle 0|\alpha_r^{(1) j_1}\, \tilde \alpha_r^{(1) j_2}\,
 \alpha_s^{(1) j_3}\, \tilde \alpha_s^{(1) j_4}
 \otimes  {}_h\langle 0|\alpha_r^{(2)p_1}\, \tilde \alpha_r^{(2)p_2}  \,
  \alpha_s^{(2) p_3}\, \tilde \alpha_s^{(2) p_4}.
\label{bostates2}
\ea
Here, we have used the property of the $D$-instanton boundary state
that the only non-zero matrix elements are those in which each
$\alpha_p$ mode is accompanied by a  $\tilde \alpha_p$ with the same
mode number, $p$.  Also, conservation of $p_+$ requires that the mode
numbers of state $(1)$ are the same as those of state $(2)$.

To leading order in the $m\to \infty$ limit the contribution to matrix
elements, ${_{{}_1}}\langle \chi_1|\otimes {_{{}_2}}\langle\chi_2|
V_2\rangle\!\rangle$,  is obtained by expanding the boundary state and
retaining the term involving the product of two fermion bilinears on
each disk.  These contributions are dominant since the fermion
bilinear in the exponent of the boundary state operator comes with an
explicit  factor of $m$ in the large-$m$ limit.  Therefore, in order
to get a non-vanishing overlap with states made of bosonic
oscillators, there must be 4 ${Q}^-$'s on each disk with one
broken supersymmetry acting on each of the fermions.  To leading order
in $m$ this leads to the following instantonic contribution to $H^{(2)}$
\ba
&&t_{j_1\tilde j_2j_3\tilde j_4}t_{p_1 \tilde p_2 p_3 \tilde
  p_4}e^{2\pi i \tau} g_s^{7/2} m^8
{1\over r^2 s^2
  }\nonumber \\
&&\int\,d^8\eta\, \eta
\gamma^{j_1}(\bbbone-\Pi)\gamma^{j_2}\eta\,\eta\gamma^{j_3}
(\bbbone-\Pi)\gamma^{j_4}\eta\,
\eta\gamma^{p_1}(\bbbone+\Pi)\gamma^{p_2}\eta\,\eta\gamma^{p_3}
(\bbbone+\Pi)\gamma^{p_4}\eta\,, \label{fourimp}
\ea
where the tilded index is associated with $\tilde\alpha$ and the
untilded with $\alpha$ and there is level matching for the first and
last pair of indices.  Considering all the indices on the external
wavefunctions to belong to the first $SO(4)$ yields for the second
line in equation (\ref{fourimp})
\be
\int\,d^8\eta\, \eta^+
\gamma^{j_1 \tilde j_2}\eta^+\,\eta^+\gamma^{j_3 \tilde j_4}\eta^+\,
\eta^-\gamma^{p_1 \tilde p_2}\eta^-\,\eta^-\gamma^{p_3 \tilde p_4}\eta^-\,,
\ee
which using equation (\ref{etaints1}) reduces to
\ba
\calT^+_{j_1\tilde j_2j_3 \tilde j_4}\calT^-_{p_1\tilde p_2p_3\tilde p_4}
&=&\bigg(\epsilon_{j_1\tilde j_2 j_3 \tilde
  j_4}+\delta_{j_1j_3}\delta_{\tilde j_2
 \tilde j_4}-\delta_{j_1 \tilde j_4}\delta_{\tilde j_2 j_3}\bigg)\nonumber \\
&\times&\bigg(\epsilon_{p_1 \tilde p_2 p_3 \tilde
  p_4}-\delta_{p_1p_3}\delta_{\tilde p_2
 \tilde  p_4}+\delta_{p_1 \tilde p_4}\delta_{\tilde p_2 p_3}\bigg)\,.
\ea
Thus, the matrix elements of two states with four bosonic impurities (with all
impurities in one of the $SO(4)$ factors) is given by
\ba
t_{j_1\tilde j_2j_3\tilde j_4}t_{p_1 \tilde p_2 p_3 \tilde p_4}e^{2\pi
  i \tau}g_s^{7/2} m^8
{1\over r^2s^2}\, \calT^+_{j_1\tilde j_2j_3\tilde
  j_4}\calT^-_{p_1\tilde p_2p_3\tilde p_4} \, .
\label{finfour}
\ea
Note that the tensors $\calT^\pm_{j_1 j_2 j_3 j_4}=
\pm\epsilon_{j_1 j_2 j_3 j_4}+\delta_{j_1j_3}\delta_{j_2
  j_4}-\delta_{j_1 j_4}\delta_{j_2 j_3}$ are self-dual, \ie,
\be
\epsilon_{j_1 j_2 p_3 p_4}\calT^\pm_{j_1 j_2 j_3 j_4}=\pm 2
\calT^\pm_{p_3 p_4 j_3 j_4}\,.
\ee
Recalling that $4\pi g_s m^2=g_2=J^2/N$, we see that the string result
predicts that the corresponding four impurity operators on the gauge
theory side should receive instanton contributions to anomalous
dimensions at order $J^7/N^{7/2}$.  The rest of the possible
four-impurity states are suppressed by powers of $m$ compared to this
leading result. For example, if the external states are of the type
${_h}\langle
0|\alpha_r^{[j_1}\tilde\alpha_r^{j_2]}\alpha_s^{[j_3}\tilde\alpha_s^{j_4]}$,
then the result will be proportional to $m^4$ rather than $m^8$. The
corresponding gauge theory result would appear at order $J^3
g_{YM}^4/N^{3/2}$.

\section{Discussion}

We have evaluated the leading single $D$-instanton contribution to
the mass matrix elements of certain states of maximally supersymmetric
plane-wave string theory.  These are contributions that are exponentially suppressed
by a factor $e^{-2\pi \tau_2}$, but are uniquely specified by
 the characteristic instanton phase,
$e^{2\pi i \tau_1}$. Our specific results concern states
with up to four impurities with non-zero mode numbers (a state
with zero mode number is a protected supergravity state and has
vanishing mass matrix element with all other states).  The
structure of the boundary state makes it obvious that the only
states that couple with a $D$-instanton are those with an even
number of impurities.   Furthermore, when there are $4 + 2n$
impurities the only states that couple to the $D$-instanton are those
in which  $n$ pairs form $SO(4)\times SO(4)$ singlets.
We also saw in section 3.2 that mass matrix elements of
  states with a large number of fermionic excitations can have a high
  positive power of $m$ (multiplying the factor of $e^{2\pi i \tau})$.
  This only arises for matrix elements which do
  not get any contributions in string perturbation theory.

It is of interest to see how the string mass matrix elements translate into
statements concerning contributions of Yang--Mills instantons to the
anomalous dimensions of states in the BMN limit of $\calN=4$
superconformal gauge theory.  
The
duality relates corrections to the masses of physical states to
anomalous
dimensions of the dual operators. A quantitative comparison requires
the
diagonalisation of both the mass matrix in string theory and the
matrix of
anomalous dimensions in the gauge theory \cite{bers}.
Since we have not diagonalized the
mass-matrix the best we can do is to compare the behaviour of individual matrix
elements.  Furthermore, in order to remain in the perturbative regime
of the string theory, $g_s\to 0$, as well as in the Yang--Mills theory,
$\lambda'\to 0$,  we have concentrated on the limit $m\to \infty$.  To
leading order as $m \to \infty$ the matrix elements between
two-impurity states given in section 3 have the form $e^{i\theta
-8\pi^2/g_{_{YM}}^2}\, g_2^{7/2}\, \lambda^{\prime 2}$, when expressed
in terms of the parameters of the Yang--Mills theory in the BMN limit.
In particular, they are independent of the mode numbers of the states.
In the case of four impurities the mass-matrix elements have the form
$e^{i\theta -8\pi^2/g_{_{YM}}^2}\, g_2^{7/2}(rs)^{-2}$,
where $r$ and $s$ are the two independent mode numbers
that label  either of the states.  In a separate paper \cite{gks} we
are examining if these dependences can be reproduced from Yang-Mills
instanton contributions to anomalous dimensions of the corresponding
four impurity operators{\footnote{After this paper was released, we
    indeed found in \cite{gks} that the coupling constant and mode
    number dependences could be reproduced on the gauge side.}}.

However, the issue of whether there should be a
precise match between the string theory and gauge theory calculations is called into
question by recent perturbative calculations.
An  analysis in \cite{gutjahr}
suggests that the one-loop string calculation in the literature is
incomplete and may be incorrect. This raises questions about the claimed
 precise match with the gauge theory analysis, which
is also incomplete.  Furthermore, perturbative calculations in
`near-BMN sectors', both in string theory \cite{nearbmnstring} and in the
dual gauge theory \cite{nearbmngauge}, have shown deviations from
BMN scaling. Explicit tests show that in the strict BMN case
scaling is respected up to three loops \cite{bmn3loop}, with
indications from a related matrix model calculation \cite{fkp} that
this may break down at four loops.
In view of these issues in the perturbative sector, it was not obvious to what extent
one should have expected agreement between the non-perturbative effects considered in this
paper  with   corresponding effects in $\calN$=4 SYM.  
An agreement in the non-perturbative sector would
suggest that
is likely that BMN scaling should hold even in the perturbative sector.

\vskip 1cm
\noindent{\bf Acknowledgments}

\noindent{\small We thank Matthias Staudacher
for useful comments. AS thanks PPARC for a post-doctoral fellowship
and Gonville and Caius college, Cambridge for a college fellowship.}
\vskip 1cm

\appendix
\label{appenone}

\section{World-sheet generators of plane-wave string theory}

In this appendix we summarise the expressions for the generators of
the plane-wave algebra in the representation furnished by string
theory in the light-cone gauge \cite{mt}.

In the light-cone gauge the 32 supersymmetries of the plane-wave background
are described by 16 kinematic and 16 dynamical supersymmetries (only the latter
receive corrections due to string interactions).
The kinematical supersymmetries, which
satisfy the anticommutation relations  $\{q,\bar q\}=p_-$,  are
given by
\ba
q&=& {e(p_-)\over 2\sqrt{2}\pi}\int_0^{2\pi|p_-|} d\sigma (S+i\tS)\equiv
\int_0^{2\pi |p_-|}d\sigma{1\over 2\sqrt{2}\pi  }\, \bar  \theta \,,\\
\bar q&=& {1\over 2\sqrt{2}\pi}\int_0^{2\pi|p_-|} d\sigma (S-i\tS)\equiv
\int_0^{2\pi|p_-|}d\sigma {e(p_-)\over 2\sqrt{2}\pi }\, \theta\,.
\label{qbardef}
\ea
The field $S$ satisfies
$\{S(\sigma),S(\sigma')\}=2\pi\delta(\sigma-\sigma')$ with a similar
relation for $\tS$. $e(p_-)={\rm sign}(p_-)$.
In (\ref{qbardef}) we have defined
\ba
e(p_-){S+i\tilde S\over 2\sqrt{2}\pi }={\bar \theta\over 2\pi
  \sqrt{2}}\,, \qquad
e(p_-){S-i\tilde S\over 2\sqrt{2}\pi }= {\theta\over 2\pi
  \sqrt{2}}\,.
\ea
The dynamical supercharges are defined as $Q^+={1\over \sqrt{2}}
(Q+i\tilde Q)$ and $Q^-={1\over \sqrt{2}}
(Q-i\tilde Q)$ with
\ba
Q^+&=&\int_0^{2\pi |p_-|}d\sigma \left[ p^I\gamma_I
\bar \theta -{i e(p_-)\over 4 \pi}\partial_\sigma x^I \gamma_I
\theta-{ie(p_-)\over 4\pi}
\mu  x^I\gamma_I \Pi \, \bar\theta \right]\,,\\
{Q}^-&=&\int_0^{2\pi|p_-|}d\sigma\left[e(p_-)p^I\gamma_I\theta+
{i\over 4\pi}\partial_\sigma x^I\gamma_I
  \bar\theta+{i\over 4\pi} \mu x^I\gamma_I \Pi\, \theta\right]\,.
\ea
In terms of these variables, the first quantised hamiltonian is
given by
\bea
h&=&{e(p_-)\over 2}\int_0^{2\pi |p_-|}d\sigma\left[
4\pi p^2+{1\over 4\pi}((\partial_\sigma x)^2+\mu^2 x^2)\right]
\nonumber \\
&+& e(p_-)\left[-4\pi \lambda\partial_\sigma \lambda+{1\over
4\pi}\theta\partial_\sigma\theta+2\mu(\lambda\Pi\theta)\right]\,.
\eea

The various mode expansions are as follows
\ba
x(\sigma,\tau=0)&=& x_0+ i\sum_{n\neq 0}{1\over
  \omega_n}\left(e^{in\sigma\over |p_-|}\alpha_n+e^{-in\sigma\over |p_-|}
  \tilde\alpha_n\right)\,, \cr
|p_-|p(\sigma,\tau=0)&=& p_0+\sum_{n\neq 0}\left(e^{in\sigma\over
  |p_-|}\alpha_n+e^{-in\sigma\over |p_-|}\tilde\alpha_n\right)\,, \cr
\sqrt{|p_-|}S(\sigma,\tau=0)&=&S_0+\sum_{n\neq 0}c_n\left[S_n
  e^{in\sigma\over |p_-|}+{i\over m}(\omega_n-n)\Pi \tilde S_n
  e^{-in\sigma\over |p_-|}\right]\,, \cr
\sqrt{|p_-|}\tilde S(\sigma,\tau=0)&=&\tilde S_0+\sum_{n\neq 0}c_n\left[\tS_n
  e^{-in\sigma\over |p_-|}-{i\over m}(\omega_n-n)\Pi S_n
  e^{in\sigma\over |p_-|}\right]\,,
\ea
where $m=\mu p_-\alpha'$ and the non-zero commutation relations are given by
\be
[\alpha_k,\alpha_l]=\omega_k \delta_{k,-l}\,,\quad
\{S_k,S_l\}=\delta_{k,-l}\,,
\ee
with similar relations for the right-movers. The quantity $c_n$ is defined by
\be
c_n ={m \over \sqrt{2\omega_n (\omega_n -n)}}\, .
\label{cndef}
\ee

\end{document}